\documentclass{aa}

\usepackage{amsbsy,graphicx,astron,epsfig}

\def \be {\begin{equation}} 
\def \en {\end{equation}} 
\def \bea {\begin{eqnarray}} 
\def \ena {\end{eqnarray}}

\def \bi{\begin{itemize}} 
\def \ei{\end{itemize}}

\def \eg {{\it e.g. }}  
\def \ie {{\it i.e. }} 
\def \etal {{\it et al. }}
\def \pr {{\it Phys. Rep.}}
\def \apj {{ \it ApJ}}
\def \apjl {{\it ApJ Let.}}
\def \pasj {{ \it PASJ}}
\def \araa {{ \it ARA\&A}}
\def \aa {{ \it A\&A}}

\begin{document}



\title{Cluster physics from joint weak gravitational lensing and Sunyaev-Zel'dovich data} 
\author{O.  Dor{\'e}\inst{1}   \and  F.R.   Bouchet\inst{1}   \and  Y.
Mellier\inst{1,2} \and R.  Teyssier\inst{3,1,4}} 
\institute{Institut d'Astrophysique de  Paris, 98bis, Boulevard Arago,
75014 Paris, FRANCE 
\and Observatoire de Paris, DEMIRM, 61 avenue de l'Observatoire, 75014
Paris, FRANCE
\and Service d'Astrophysique, DAPNIA, Centre d'{\'E}tudes de Saclay, 91191
Gif-sur-Yvette, FRANCE
\and Numerical Investigations in Cosmology (NIC) group, CEA Saclay }

\offprints{O.~Dor{\'e}} 
\mail{dore@iap.fr} 
\markboth{Cluster physics from joint WL and SZ data}{Dor{\'e} \etal} 
\date{}

\abstract
{
We present a self consistent  method to perfom a joint analysis  of
Sunyaev-Zel'dovich and weak gravitational lensing observation of galaxy
clusters.  The spatial distribution of the cluster main constituents  is
described by a perturbative approach. Assuming the hydrostatic equilibrium and the
equation of state, we are able to deduce, from observations, maps of
projected gas density and gas temperature. The method then naturally
entails an X-ray  emissivity prediction which can be compared to
observed X-ray emissivity maps. When tested on simulated clusters
(noise free), this prediction turns out to be in very good agreement
with the simulated surface brightness. The simulated and predicted
surface brightness images have a correlation coefficient higher than $0.9$ and the
total flux   differ by $0.9 \%$ or $9 \% $ in the two simulated clusters we
studied. The method should be easily used on real data in order to provide a
physical description of the cluster physics and of its constituents.  The tests
performed show that we can recover  the amount and the spatial distributions of both
the baryonic and non-baryonic material with an accuracy  better than
10\%.  So, in principle, in it might indeed help to alleviate some
well known bias affecting, \eg baryon fraction measurements.}

\maketitle


\section{Introduction}

Whereas  clusters of  galaxies, as the largest gravitationnaly bound
structures of  the universe, form natural probe of cosmology,
observations, numerical simulations as well as timing arguments
provide compelling evidences that most of them are young and complex
systems. Interaction with large-scale structures, merging processes
and coupling of dark matter with the intra-cluster medium complicate
the interpretation of observations and the modeling of each of its
components. Since they are composed of  dark mater (DM),  galaxies and
a hot dilute X-ray emitting gas  (Intra cluster medium, ICM) accounting respectively  for
$\sim 85\%$, $\sim 15\%$ and $\sim 5\%$  of their mass,  the physics
of  the ICM bounded in  a dark matter gravitational potential  plays a
major role in cluster  formation and  evolution.  This  variety of
components  can  be observed  in  many  various  ways. In  particular,
gravitational  lensing  effects  (the  weak-lensing regime  here,  WL)
\cite{Me00,BaSc01}, Sunyaev-Zel'dovich  (SZ) effect \cite{SuZe72,Bi99}
and X-ray emission (X) \cite{Sa88}. Whereas the  former probes mostly
the dark matter component, both the latter probe the baryons  of the
gravitationally bound ICM. \\

Due to observational progress, increasingly high quality data are
delivered which enables multi-wavelength investigation of clusters on
arcminute scale (the most recent is the spectacular progress in SZ
measurements, \eg \cite{ReMo20,DeBe98}) and we therefore think it is
timely to explore how we should perform some joint analysis of
these high quality data sets and exploit them at best their
complementarity. This challenge has already been tackled  by several groups
\cite{ZaSq98,GrCa99,Re00,ZaSq00,Ca00,Ho00}. Zaroubi \etal and
Reblinsky \etal attempted a full deprojection by assuming
isothermality and axial symmetry, using respectively a least square
minimization or a Lucy-Richardson algorithm , Grego \etal compare SZ
derived  gas mass to WL derived total mass by fitting a spheroidal
$\beta$ model. But whereas these methods give reasonable results it
has been illustrated, \eg by Inagaki \etal 1995 in the context of
$H_0$ measurement from SZ and X-ray observations,  that both non 
isothermality and asphericity analysis can trigger systematic errors
as high as $20\ \%$. Therefore, we aim at exploring an original
approach which allows to get rid of both isothermality and departure
from sphericity. Based on a self-consistent use of both observables,
and based on a perturbative development of general physical
hypothesis, this method allow us to test some very general
physical  hypothesis  of  the  gas  (hydrostatic  equilibrium,  global
thermodynamic equilibrium) and also provide naturally some X
observation predictions.\\

Observations only provide us with $2-D$ projected quantities (\eg mass, gas
pression,\ldots).   This  quantities  are  related  by  some  physical
hypothesis which  are explicited in $3-D$  equalities (\eg hydrostatic
equilibrium,  equation  of  state).  The  point is  that  these  $3-D$
equalities  do not  have any  tractable equivalent  relating projected
$2-D$ quantities:  in particular, projection  along the line  of sight
does  not provide  an equation  of  state or  a projected  hydrostatic
equilibrium equation.  Therefore as  soon as we  want to  compare this
data (WL, SZ, X) we have  to deproject the relevant physical
quantities ($P_g, T_g,  \rho_g$\ldots). This can be done only
using  strong assumptions, either  by using  parametric models  (\eg a
$\beta$   model  \cite{CaFu76})  or   by  assuming   mere  geometrical
hypothesis   (the   former   necessarily  encompassing   the   latter)
\cite{FaHu81,YoSu99}.  We choose the  geometric approach in  order to
use  as general  physical grounds  as possible  and to  avoid  as many
theoretical biases as possible.\\

This simplest choice  might be naturally motivated first  by looking at
some images of observed clusters \cite{DeBe98,GrCa99}. Their regularity is
striking : some  have almost circular or ellipsoidal  appearance as we
expect  for fully  relaxed  system. Then  since  relaxed clusters  are
expected to be spheroidal  in favored hierarchical structure formation
scenario,   it   is   natural   to   try  to   relate   the   observed
quasi-circularity  (quasi-sphericity)  to  the $3-D$  quasi-sphericity
(quasi-spheroidality). We  perform this using  some linearly perturbed
spherical (spheroidal) symmetries in a self-consistent approach. \\

We  proceed as  follows:  in  section \ref{notation}  we defined  our
physical  hypothesis  and  our  notations.  The  method  is  precisely
described in  section \ref{method}. We consider both  the spherical as
well as spheroidal  cases and obtain a predicted  X surface brightness
map from a SZ decrement map  and a WL gravitational distortion map. In
section  \ref{simulation} a demonstration  with simulated  clusters is
presented before discussing its application to genuine data as well as
further developments in section \ref{discussion}.

\section{Hypothesis, Sunyaev-Zel'dovich effect and the Weak lensing}\label{notation}

We  now  briefly  describe  our  notations as  well  as  our  physical
hypothesis.

\subsection{General hypothesis}
\label{hypo_gen}
Following considerations fully detailed  in \cite{Sa88} the ICM can be
regarded as  a hot and  dilute plasma constituted  from ions and
electrons, whose respective kinetic temperatures $T_p$ and $T_e$  will
be considered as equal $T_{p}=T_{e} \equiv T_{g}$. This is the \emph{global thermodynamic
equilibrium hypothesis} which is expected to hold up to $r_{virial}$ (
see  \cite{TeCh97,ChAl98} for  a  precise discussion).  Given the  low
density  (from $n_e\sim  10^{-1} \rm{cm^{-3}}$  in the  core  to $\sim
10^{-5} \rm{cm^{-3}}$ in the outer  part) and high temperature of this
plasma  ($\sim 10  \rm{keV}$),  it can  be  treated as  a perfect  gas
satisfying the equation of state : 
\be 
P_{g}\ = \ \frac{\rho_{g}\ k_{B}\ T_{g}}{\mu_{e} \ m_{p}} \ = \ \beta\ \rho_{g}T_{g}
\label{state} 
\en 
with   $\beta   \equiv  \frac{\   k_{B}}{\mu_{e}\
m_{p}}$. Let us neglect then the gas mass with regards to the dark matter
mass, and assume \emph{stationarity} (no   gravitational  potential
variation on time scale smaller  than the hydrodynamic time scale, \eg
no  recent  mergers). Then the gas assumed to be in hydrostatic equilibrium in the
dark matter gravitational potential satisfies:
\begin{eqnarray}
\nabla(\rho_g\mathbf{v_g})  &=& 0  \\ \nabla  P_{g}  &=& -\rho_g\nabla
\Phi_{DM}\: .
\label{hydrostat2}
\end{eqnarray} 
At this point there is no need to assume isothermality.\\

\subsection{Sunyaev-Zel'dovich effect and weak lensing}

Inverse Compton  scattering of cosmic background (CMB)  photons by the
electrons     in    the    ICM     modifies    the     CMB    spectrum
\cite{ZeSu69,SuZe72,SuZe80a}.   The amplitude  of  the SZ  temperature
decrement {$\Delta T_{SZ} \over  T_{CMB}$} is directly proportional to
the Comptonisation parameter $y$ which is given by : 
\bea 
y\ =\ \frac{\sigma_{T}}{m_{e}c^{2}}\int dl\ n_e k_B T_e \ & = & \
\frac{\sigma_{T}}{m_{e}c^{2}}\int dl\ p_e \\ 
=\ \frac{\sigma_{T}}{m_{e}c^{2}}\int dl \frac{\rho_g k_B T_g}{\mu_e m_p}\
& = & \ \alpha \int dl P_g\: .  
\ena 
where $\alpha \equiv \frac{\sigma_{T}}{m_{e}c^{2}}$, $k_B$ is the Boltzmann's constant,
$\sigma_{T}$ is the  Thomson scattering cross section and  $dl$ is the
physical line-of-sight  distance.  $m_e$,  $n_e$, $T_e$ and  $p_e$ are
the mass, the number density, the temperature and the thermal pressure
of electrons. $\rho_g $ and  $T_g$ respectively denote the gas density
and temperature,  and $\mu_e$  is the number  of electrons  per proton
mass.  Some further  corrections to  this expression  can be  found in
\cite{Re95,Bi99}. \\

In   parallel   to   this   spectral  distortions,   the   statistical
determination  of the  shear field  $\kappa$ affecting  the  images of
background galaxies enable, in the  weak lensing regime, to derive the
dominant projected gravitational potential  of the lens (the clustered
dark matter) : $\phi_{DM}$ in our general hypothesis (see \cite{Me00} for details).

\section{Method}\label{method}

\subsection{Principle}

We now  answer the question :  how should we  co-analyze these various
data set ?   Our first aim is  to develop a method which  allows us to
get maps  of projected  thermodynamical quantities  with  as few
physical hypothesis as possible.\\

\begin{figure}[!tb] 
\begin{center}
\psfig{file=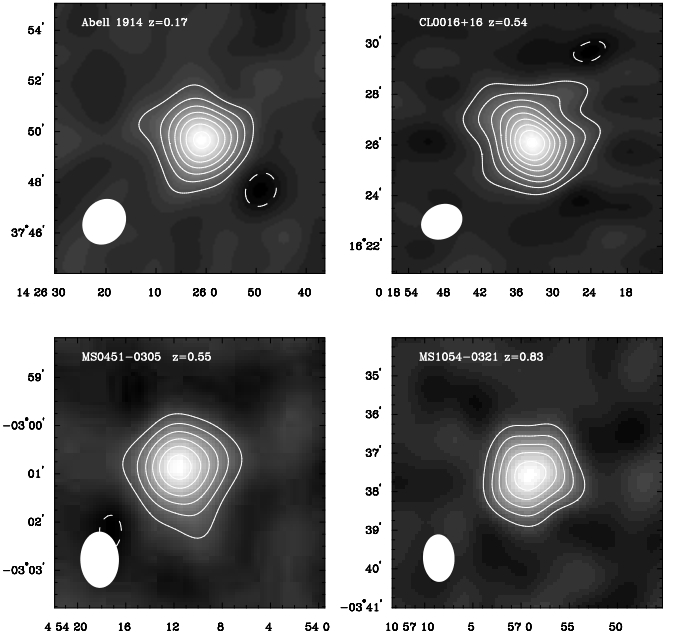, width= \hsize, angle=0}
\end{center}
\caption{Images of the SZ effect observed towards four galaxy clusters
with various redshifts. The contours  correspond to 1.5 to 5 times the
noise level. Data taken  with the low-noise cm-wave receiver installed
on  the OVRO  and  BIMA mm-wave  interferometric  arrays (Holder  and
Carlstrom 1999).\label{szfig}}
\end{figure}

Our  method is  the following.  Let  us suppose  we have  for a  given
cluster a set of data a SZ and WL data which enables us to construct a
$2-D$  map of  projected gas  pressure as  well as  a  $2-D$ projected
gravitational potential  map. Let us  suppose as well that  these maps
exhibit an approximate spherical symmetry as it is the case for a vast
class of experimental observations  as \eg in figure \ref{szfig}. More
precisely, let us suppose that  the projected gas pressure $y$ as well
as the  observed projected gravitational potential  $\phi_{DM}$ can be
well fitted by the following type of functions~: 
\bea 
y(R, \varphi) & = & y_0(R) + \varepsilon y_1(R)\ m(\varphi) \\ 
\phi_{DM}(R,   \varphi)   &   =   &   \phi_{DM,0}(R)   +   \varepsilon
\phi_{DM,1}(R)\ n(\varphi) 
\ena 
where $\varepsilon \ll 1$, $(R,\varphi)$ denotes polar coordinates in the
image  plane and  $m$  and  $n$ are  some  particular functions.  This
description  means first  of all  that the  images we  see  are linear
perturbations  from  some  perfect  circularly symmetric  images,  and
second that  the perturbation might  be described conveniently  by the
product of a radial function  and an angular function. Equivalently we
can  assert  that to  first  order  in  $\varepsilon$ our  images  are
circularly symmetric  but they admit some corrections  to second order
in  $\varepsilon$.\\
We  then  assume that  these  observed  perturbed
symmetries are a consequence  of an intrinsic $3-D$ spherical symmetry
linearly perturbed  too.  This  point constitutes our  key hypothesis.
It  means   that  to   first  order  in   a  certain   parameter  (\eg
$\varepsilon$) our clusters are regular objects with a strong circular
symmetry but  they admit some  second order linear  perturbations away
from this symmetry. As a consequence of these assumptions we will make
use  of  this  linearly  perturbed  symmetry  to get  a  map  of  some
complementary  projected thermodynamical  quantities, the  gas density
$D_g$  and the gas  temperature $\zeta_g$,  successively to  first and
second order in $\varepsilon$.\\

Formulated this way, the problem yields a natural protocol~:
\begin{itemize} 
\item Looking  at some maps with  this kind of symmetry,  we compute a
zero-order  map  ($y_0(R)$,   $\phi_0(R)$)  with  a  perfect  circular
symmetry by  averaging over some concentric annulus.  A correction for
the bias  introduced by perturbations  is included. These  first order
quantities allow  us to derive  some first order maps  of $D_{g,0}(R)$
and $\zeta_{g,0}(R)$ with a perfect circular symmetry.
\item We  then take into account  the first order  corrections to this
perfect symmetry ($y_1(R)m(\varphi)$, $\phi_1(R)m(\varphi)$) and infer
from them first order correction terms to the zeroth order maps:
$D_{g,1}(R,\varphi)$ and $\zeta_{g,1}(R,\varphi)$.
\end{itemize}   

Even if for  clarity's sake we formulate our  method  assuming  a
perturbed circular  symmetry,  it  applies equivalently  to a
perturbed elliptical symmetry  as it will be  shown below. In this more
general case, we assume that the cluster exhibit a linearly perturbed spheroidal symmetry.\\

\subsection{The spherically symmetric case : from observations to predictions}

Let  us now  apply the  method  to the  case where  the projected  gas
density (SZ data) and  the projected gravitational potential (WL data)
exhibit some  approximate circular symmetry.  These  observations lead
us to suppose  that the  $3-D$ gas pressure, the gravitational
potential, the gas  density and  the gas temperature can be well
described by the following equations:
\be
\left\{
\begin{array}{cccccc}
P_g(r,\theta,\varphi) &=& P_{g,0}(r) &+& \varepsilon & P_{g,1}(r)f(\theta,\varphi) \\   
\Phi_{DM}(r, \theta,\varphi) &=& \Phi_{DM,0}(r) &+&  \varepsilon &
\Phi_{DM,1}(r)g(\theta,\varphi) \\
\rho_g(r,\theta,\varphi) &=& \rho_{g,0}(r) &+& \varepsilon &
\rho_{g,1}(r)h(\theta,\varphi) \\ 
T_g(r,\theta,\varphi) &=& T_{g,0}(r) &+& \varepsilon & T_{g,1}(r)k(\theta,\varphi)
\end{array}
\right.
\en
where $(r,\theta,\varphi)$  are spherical coordinates  centered on the
cluster.

\subsubsection{The hydrostatic equilibrium}
If we first  apply the hydrostatic equilibrium equation  $\nabla P_g =
-\rho_g \nabla  \Phi_{DM}$ we get  the following equations.   To first
order in $\varepsilon$ we have
\begin{equation}
P'_{g,0}(r)\ =\ -\rho_{g,0}(r)\Phi'_{DM,0}(r)\: ,
\label{hydro1e}
\end{equation}
and to second order in $\varepsilon$ :
\begin{equation}
\left\{
\begin{array}{cccc}
P'_{g,1}(r)f(\theta,\varphi)\    &=&\   -\rho_{g,0}(r)\Phi'_{DM,1}(r)\
h(\theta,\varphi) & \\ 
& & - \rho_{g,1}(r)\Phi'_{DM,0}(r)\ g(\theta,\varphi) &\: (a)\\ 
P_{g,1}(r)\ \partial_{\theta}\ f(\theta,\varphi)\ &=&\ -\rho_{g,0}(r)\ \Phi_{DM,1}(r)\
\partial_{\theta} h(\theta,\varphi) &\: (b) \\ 
P_{g,1}(r)\ \partial_{\varphi} f(\theta,\varphi)\ &=&\ - \rho_{g,0}(r)\
\Phi_{DM,1}(r)\ \partial_{\varphi} h(\theta,\varphi) &\: (c)
\end{array}\right.
\label{hydro2e}
\end{equation}
where  \  `` '  ``\  denotes the  derivative  with  regards to  $r$.\\
Combining  equations (\ref{hydro2e}.b)  and  (\ref{hydro2e}.c) we  get
\bea 
f(\theta,\varphi) = \lambda_1 h(\theta,\varphi) + \lambda_2 
\ena
where  $\lambda_{1,2}$   are  some  constants. Then, using equation
(\ref{hydro2e}.a) we can write 
\bea 
f(\theta,\varphi)  =  \gamma_1
g(\theta,\varphi) + \gamma_2 
\ena 
where $\gamma_{1,2}$ are some constants as well. At this point, we can
get rid of  $\lambda_2$ and $\gamma_2$ by absorbing  them in the order
$1$ mere radial term (\ie $\rho_{g,0}(r)$ and $\Phi_{DM,0}(r)$). This means we can consider
$\lambda_2=0$  and  $\gamma_2=0$  .  Similarly we  choose  to  rescale
$\rho_{g,1}(r)$ and  $\Phi_{DM,1}(r)$ so that we can  take $\gamma_1 =
\lambda_1 = 1\ $.  These simple  equalities lead us to assume from now
on : 
\be 
f(\theta,\varphi) = h(\theta,\varphi) = g(\theta,\varphi)\:.
\label{fgh}
\en 
This is in no  way a restriction since  it simply means
that we absorb integration constants by redefining some terms. This is
possible since the relevant part of  $f$ (and thus $h$) will be fitted
on observations  as will be shown below.   Taking equation (\ref{fgh})
into account, equation (\ref{hydro2e}) simplifies to : 
\bea
P'_{g,0}(r)\ &=&\ -\rho_{g,0}(r)\Phi'_{DM,0}(r)\label{hydrosimple1}\\
P'_{g,1}(r)\ &=&\ -\rho_{g,0}(r)\Phi'_{DM,1}(r) - \rho_{g,1}(r)\Phi'_{DM,0}(r) 
\label{hydrosimple2}\\ 
P_{g,1}(r)\ &=&\ -\rho_{g,0}(r)\Phi_{DM,1}(r) \: .
\label{hydrosimple3}
\ena

\subsubsection{The equation of state}

We have now identified the  angular part to the first order correction
of  $P_g$, $\Phi_{DM}$  and  $\rho_g$.  We still  have  to link  those
quantities  to the angular  dependent part  of the  temperature $T_g$,
namely $k(\theta,\varphi)$. This is  done naturally using the equation
of  state (\ref{state}), which  directly provide  to first  and second
order in $\varepsilon$ : 
\bea 
P_{g,0}(r)\ &=&\ \beta \rho_{g,0}(r) T_{g,0}(r) \label{state0}\\
P_{g,1}(r)  f(\theta,\varphi) \ &=&\ \beta
\rho_{g,1}(r) T_{g,0}(r)f(\theta,\varphi) \nonumber \\ 
\;   &    +   &   \beta    \rho_{g,0}(r)   T_{g,1}(r)k(\theta,\varphi)
\label{state1} 
\ena
This last equation leads naturally to
$f(\theta,\varphi)=k(\theta,\varphi)$  if  we decide once again to
absorb any multiplicative factor in the radial part. This way we see
that our choice of separating the radial and angular part is in no way
a restriction.  We eventually get 
\bea 
P_{g,0}(r)\ &=&\ \beta \rho_{g,0}(r) T_{g,0}(r) \\
P_{g,1}(r) \ &=&\ \beta \rho_{g,1}(r) T_{g,0}(r) + \beta \rho_{g,0}(r)
T_{g,1}(r) \: . 
\ena

\subsubsection{The observations}
\label{observations}
Given this description of the cluster hot gas, the experimental SZ and
WL data  which respectively provide  us with the  projected quantities
$y(R,\varphi)$ and $\phi_{DM}(R,\varphi)$ write 
\bea 
y(R,\varphi) & = & \alpha \int P_{g,0}(r)dl + \varepsilon\ \alpha \int
P_{g,1}(r)f(\theta,\varphi) dl\ \nonumber \\ 
& \equiv & \ y_0(R) + \varepsilon y_1(R)m(\varphi) \\
\phi_{DM}(R,\varphi) & = & \int \Phi_{DM,0}(r)dl + \varepsilon \int \Phi_{DM,1}(r)f(\theta,\varphi)
dl\ \nonumber \\ 
& \equiv & \ \phi_{DM,0}(R) + \varepsilon \phi_{DM,1}(R)m(\varphi)\: .  
\ena 
Note that in order to get  this set of definitions we choose the polar
axis of the cluster along the line of sight so that the same azimuthal
angle $\varphi$ is used for $2-D$ and $3-D$ quantities. 

Our  aim  is now  to  derive  both a  projected  gas  density map  and
projected   temperature  map   that  we   define  this   way   :  
\bea
D_g(R,\varphi)  &  =  & \int\  \rho_g(r,\varphi)  dl  \\  
& =  &  \int \rho_{g,0}(r)dl + \varepsilon \int \rho_{g,1}(r)f(\theta,\varphi) dl\\
&  \equiv  &  \   D_{g,0}(R)  +  D_{g,1}(R,  \varphi)  \label{dgdef}\\
\zeta_g(R,\varphi)  & =  &  \int\ T_g(r,\varphi)  dl  \\ 
&  = &  \int\ T_{g,0}(r)\ dl + \varepsilon \int\
T_{g,1}(r)f(\theta,\varphi) dl \\ 
& \equiv & \zeta_{g,0}(R)  + \zeta_{g,1}(R,\varphi)\: . \label{zetagdef}
\ena

\subsubsection{A projected gas density map to first order\ldots} 

Now that we have expressed  our observables in terms of $3-D$ physical
quantities,  it is easy  to infer  a gas  density map  successively to
first  and  second  order   in  $\varepsilon$.   To  first  order  the
hydrostatic  equilibrium  condition  (\ref{hydro1e}) states  that  \be
P'_{g,0}(r)\ =\ -\rho_{g,0}(r)\Phi'_{DM,0}(r)\: .  \en In order to use
it we need to deproject  the relevant quantities.  From the well known
spherical deprojection formula \cite{BiTr87} based on Abel's transform
we have :
\begin{eqnarray}
\alpha\  P_{g,0}(r)  &  =  & -{1\over  \pi}\int_r^{\infty}  y'_0(R){dR
     \over(R^2-r^2)^{1\over2}} \\ & = & -{1\over \pi} \int_0^{\infty}\
     y'_0(r\cosh u)du
\label{pg0}
\end{eqnarray}
where $ R=r\cosh u$. Thus, we can write
\begin{eqnarray}
\alpha\  P'_{g,0}(r) &  = &  -{1\over \pi}  \int_0^{\infty}\  \cosh u\
      y''(r\cosh u)du \\ &  = & -{1\over \pi} \int_r^{\infty}\ {1\over
      r}{R\over (R^2-r^2)^{1 \over 2}}\ y''_0(R) dR \; .
\label{pg'0}
\end{eqnarray}
Similarly,
\begin{equation} 
\Phi'_{DM,0}(r)  = - {1\over  \pi} \int_r^{\infty}\  {1\over r}{R\over
      (R^2-r^2)^{1 \over 2}}\ \phi_0'' (R) dR \; .
\end{equation}
We then get for the  projected gas density \bea \lefteqn{ D_{g,0}(R) =
-2\int_R^{\infty}{r                                             dr\over
(r^2-R^2)^{1\over2}}{P'_{g,0}(r)\over\Phi'_{DM,0}(r)} }  \\ & &  = -{2
\over   \alpha}   \int_R^{\infty}   {r  dr\over   (r^2-R^2)^{1\over2}}
\left(\frac{   \int_r^{\infty}{   s\  ds\over   r(s^2-r^2)^{1\over2}}\
y''_0(s)}   {\int_r^{\infty}{    s\   ds\over   r(s^2-r^2)^{1\over2}}\
\phi''_0(s)} \right) \: .  \ena

\subsubsection{\ldots and a projected gas temperature map to first order}
Once  we built  this projected  gas density  map, we  can  recover the
projected  gas temperature  map. If  we  apply the  equation of  state
(\ref{state0}) we get :
\begin{eqnarray}
\zeta_{g,0}(R)   &  =   &  {1\over   \beta}  \int   {P_{g,0}(r)  \over
         \rho_{g,0}(r)}dl   \\   &  =   &   -{1\over   \beta  }   \int
         {P_{g,0}(r)\over  P'_{g,0}(r)}\Phi'_{DM,0}(r)  dl  \\ &  =  &
         -{1\over    \pi\beta}    \int_R^{\infty}\    {P_{g,0}(r)\over
         P'_{g,0}(r)}\Phi'_{DM,0}(r){rdr\over (r^2-R^2)^{1 \over 2}}\:
         .
\end{eqnarray}

Since   all    the   required   functions    ($P_{g,0}$,   $P'_{g,0}$,
$\Phi'_{DM,0}$) have  been derived  in the previous  section (equation
(\ref{pg0})  and (\ref{pg'0}))  we can  get this  way a  projected gas
temperature map.

\subsubsection{Corrections from departure to spherical symmetry~: a
projected gas density map to second order\ldots}

We now  reach the core  of our method,  namely we aim at  deriving the
quantity  $D_{g,1}$ defined  by  (\ref{dgdef}), \ie  the second  order
correction to the perfectly circular  term : \bea D_g(R,\varphi) & = &
\  D_{g,0}(R)  +  \varepsilon  D_{g,1}(R,  \varphi)  \\  &  =  &  \int
\rho_{g,0}(r)dl + \varepsilon \int \rho_{g,1}(r)f(\theta,\varphi) dl\:
.  \ena If we derive equation (\ref{hydrosimple3}) and combine it with
equation  (\ref{hydrosimple2})   we  note  that   \bea  \rho'_{g,0}(r)
\Phi_{DM,1}(r)\ &=&\ \rho_{g,1}(r) \Phi'_{DM,0}(r)\:.
\label{approxrho1}
\ena Therefore we can write \bea \int\rho_{g,1}(r)f(\theta,\varphi) dl
=         \int        \        \frac{\rho'_{g,0}(r)}{\Phi'_{DM,0}(r)}\
\Phi_{DM,1}(r)f(\theta,\varphi) dl\: .
\label{intd1g}
\ena At this point we want to express this quantity either in terms of
WL data or  in terms of SZ  data depending on the quality  of them, or
even better in terms of an optimal combination of them.\\

On one hand, WL data provide us with a straightforward access to the function
$\phi_1(R)m(\varphi)\ =\ \int \Phi_{DM,1}(r)f(\theta,\varphi) dl$ thus
we choose to approximate (\ref{intd1g}) by 
\bea
\lefteqn{\int\rho_{g,1}(r)f(\theta,\varphi) dl \simeq
\frac{\rho'_{g,0}(R)}{\Phi'_{DM,0}(R)} \int \Phi_{DM,1}(r)
f(\theta,\varphi) dl} \nonumber\\ 
& &\: \simeq \frac{\rho'_{g,0}(R)}{\Phi'_{DM,0}(R)} \phi_1(R)m(\varphi) 
\nonumber\\
& &\: \simeq \frac{\rho'_{g,0}(R)}{\Phi'_{DM,0}(R)}
\left(\phi_{DM}(R,\varphi) - \phi_0(R)\right)
\label{rho1phi}
\ena 
where we used the definitions of section (\ref{observations}) and
where $R$ corresponds  to the radius observed in  the image plane, \ie
the radius $r$ equal to the distance between the line of sight and the
center of the cluster. We will discuss this approximation in
more  details in  section  (\ref{approx}) and  validate  it through  a
practical implementation on simulations in section
(\ref{simulation}). But we already can make the following statements:
would the line of sight follows a line of  constant  $r$ throughout
the  domain  of  the perturbation,  this expression would be
rigorously exact. Moreover it turns out to be a good approximation
because of the finite extent of the perturbation.\\

On the other hand SZ data provide us with a measurement of  the function
$y_1(R)m(\varphi)\  =\ \int P_{g,1}(r)f(\theta,\varphi)  dl$ therefore
we can  use equation (\ref{hydrosimple3})  and (\ref{hydrosimple1}) to
write  
\bea  
\lefteqn{\int\rho_{g,1}(r)f(\theta,\varphi) dl = \int
\frac{\rho'_{g,0}(r)} {P'_{g,0}(r)}\  P_{g,1}(r)f(\theta,\varphi)dl}  \\ 
&\simeq& \frac{\rho'_{g,0}(R)}{P'_{g,0}(R)} \int P_{g,1}(r)f(\theta,\varphi)  dl\\ 
&\simeq& \frac{\rho'_{g,0}(R)}{P'_{g,0}(R)} y_1(R)m(\varphi)\\
&\simeq& \frac{\rho'_{g,0}(R)}{P'_{g,0}(R)} \left(y(R,\varphi) -
y_0(R)\right) \ . 
\label{rho1y}
\ena
Here again we  used the  same notation  and approximation  as in
equation (\ref{rho1phi}). Note however that as soon as we assumed
isothermality, the ratio $\rho'_{g,0}/P'_{g,0} \displaystyle$ is
constant therefore this last step is exact. Were we not assuming
isothermality, the departure from isothermality is expected to be weak
thus this last approximation should be reasonable.

This last two alternative steps  are crucial to our method since these
approximations  link  the  non  spherically  symmetric  components  of
various  quantities.  They  are  reasonable as  will  be discussed  in
section  (\ref{approx})  and will  be  numerically  tested in  section
(\ref{simulation}).\\ 
Of course,  only well-known quantities appear in equation
(\ref{rho1phi}) and  (\ref{rho1y}): $y$,  $y_0$, $\phi_{DM}$
and $\phi_0$  are direct  observational data whereas  $P_{g,0}(r)$ and
$\rho_{g,0}(r)$ are zeroth order quantities previously derived.

\subsubsection{\ldots and a projected gas temperature map to second order}

The projected temperature  map can be obtained the same way as before.
Using first the equation  of state we can  write : 
\bea
\lefteqn{T_{g,0}(r)   +\varepsilon   T_{g,1}(r)  f(\theta,\varphi)   =
{1\over     \beta}{    (P_{g,0}(r)     +     \varepsilon    P_{g,1}(r)
f(\theta,\varphi))  \over  (\rho_{g,0}(r)  +\varepsilon  \rho_{g,1}(r)
f(\theta,\varphi))}}\nonumber\\      
& \simeq & {1\over\beta}\left({P_{g,0}(r)\over\rho_{g,0}(r)} + \varepsilon P_{g,1}(r)
{\rho_{g,0}(r)  - \rho_{g,1}(r)\over\rho^2_{g,0}(r)} f(\theta,\varphi)
\right)\;.    
\ena  
Hence,   since  
\bea   
\zeta(R,\varphi)  &   =  &\ \zeta_0(R,\varphi)  + \varepsilon  \zeta_1(R,\varphi) \\  
& =  &\ \int T_{g,0}(r)\ dl +  \varepsilon \int T_{g,1}(r)f(\theta,\varphi) dl 
\ena
we   have   
\bea    
\zeta_1(R,\varphi)   =   \int   {\rho_{g,0}(r)   -
\rho_{g,1}(r)\over\rho^2_{g,0}(r)}\  P_{g,1}(r) f(\theta,\varphi) dl\:.  
\ena 
Here we choose  to approximate the last integral as previously
discussed in order to make  use of observational SZ data. Therefore we
rewrite  this  last  equation  as :  
\bea  
\lefteqn{\zeta_1(R,\varphi) \simeq  {\rho_{g,0}(R)  -  \rho_{g,1}(R)\over  \rho^2_{g,0}(R)}\  \int
P_{g,1}(r)f(\theta,\varphi) dl} \nonumber  \\ 
&\simeq & {\rho_{g,0}(R) - \rho_{g,1}(R) \over \rho^2_{g,0}(R)}\
y_1(R)m(\varphi) \\ 
&\simeq & {\rho_{g,0}(R) - \rho_{g,1}(R) \over \rho^2_{g,0}(R)}
\left(y(R,\varphi)  - y_0(R)\right)\:.   
\ena  
We obtain  this way  an expression to second  order for the projected
temperature in terms of either observed quantities or previously derived functions. \\

\subsubsection{Why the previous approximation is reasonable on intuitive grounds?}
\label{approx}
\begin{figure}[!tb] 
\begin{center}
\psfig{file = 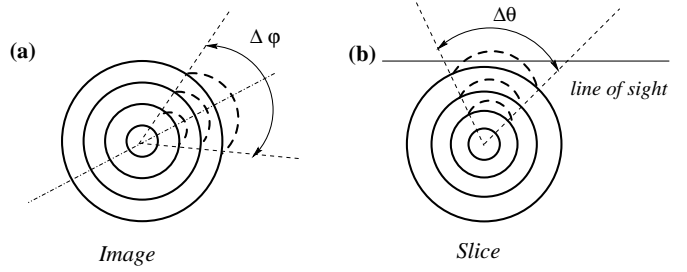, angle=0, width= \hsize}
\end{center}
\caption{We represent  schematically in (a) an  image corresponding to
our hypothesis.  The full line  corresponds to the  perfectly circular
$2-D$  term, \eg  $\phi_{DM,0}$,  and  the dashed  line  to the  first
perturbative  correction to  it, \eg  $\phi_{DM,1}m(\varphi)$, $\Delta
\varphi$ represents the observed angular extent. In (b) we represent a
schematic  slice   in  the   $3-D$  potential  responsible   for  this
image. This  slice has been performed along  the dash-two-dotted plane
indicated on figure (a). Here  again, the full line corresponds to the
perfectly circular $3-D$ term,  \eg $\Phi_{DM,0}$, and the dashed line
to    the     first    perturbative    correction     to    it,    \eg
$\Phi_{DM,1}f(\theta,\varphi)$.  The   line  of  sight   direction  is
indicated by the full thin  line. Were the line of sight perpendicular
to  this slice  plane, we  would  observe the  angular extent  $\Delta
\theta$. Giving  an axial  symmetry to this  perturbation leads  us to
assess that $\Delta \varphi \simeq \Delta \theta$.\label{fig_approx}}
\end{figure}

Our previous approximations can be justified on intuitive grounds even
if  we  will  take  care  of  validating  it  numerically  in  section
(\ref{simulation})  below. It  relies on  the fact  that perturbations
have by definition a finite  extent, \ie the first order correction to
the  perfectly circular  (spherical) term is non zero only within a finite
range. The typical size and the amplitude of the perturbation can be
easily scaled from the SZ and WL data set. This guarantees the
validity of our assumptions on observational grounds. The key point
is that the perturbation itself has a kind of axial symmetry, whose axis goes
through the center of the cluster and the peak of the
perturbation. This is reasonable if the perturbation originates in \eg
an  incoming filament  but not  for a  substructure. The  latter would
therefore have to be  treated separately by superposition (see section
(\ref{discussion})).  This  leads naturally to the  statement that the
typical angle  we observe in  the image plane  is equal to the  one we
would observe  if the line of  sight were perpendicular  to its actual
direction,  \ie the  perturbation  as intrinsically  the same  angular
extent in the directions along  the line of sight and perpendicular to
it.  This is illustrated schematically in figure (\ref{fig_approx}).

\begin{figure}[!tb]
\begin{center}
\psfig{file=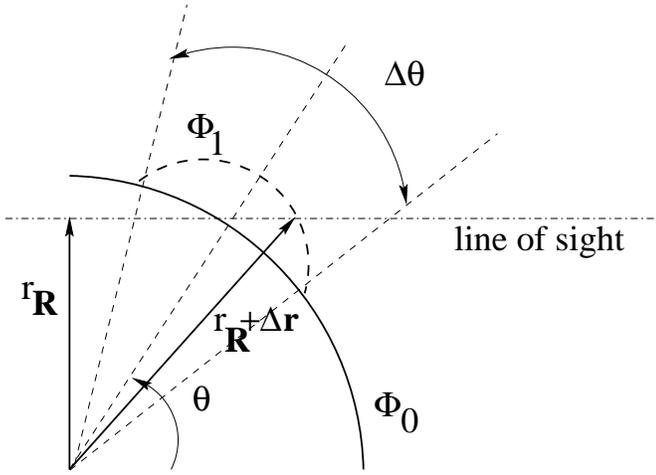,width= \hsize, angle=0}
\end{center}
\caption{We define  in this figure  the notation necessary  to discuss
our approximation. $r_R$ is the  parameter value given to the function
$(f(r))_R$ and  $r_R+\Delta r$  is an actual  value along the  line of
sight.\label{fig_approx2}}
\end{figure}

Given  this  description we  are  now in  a  position  to discuss  the
validity of  our approximation. It  consists in approximating  the
line  of sight  integral $\int  g(r)\Phi_{DM,1}(r)f(\theta,\varphi) dl
\displaystyle$   by   $g(R)   \int   \Phi_{DM,1}(r)f(\theta,\varphi)
dl\displaystyle$ where $g$ is  any radial function. This approximation
would be exact if $g(r)$ were  constant in the relevant domain, \ie if
the line of sight had a  constant $r$. As mentioned before this is the
case in equation \ref{rho1y} if we assume isothermality. But the
functions $g(r)$  we might deal with may scale roughly as $r^2$, as
\eg $\rho'_{g,0}(r)/P_{g,0}(r)$ in equation  (\ref{rho1phi}), thus  it is
far  from  being constant.  The  consequent  error  committed can  be
estimated by the quantity $\Delta r g'(r)$ where $\Delta r$ is
the maximum  $r$ discrepancy between  the value assumed, $g(R)$, and
the  actual  value  as  it  is  schematically  illustrated  in  figure
(\ref{fig_approx2}).  In the worst case, $g'(r)$  scales  as $r$. Then, using  the
obvious notations  defined in this  figure we get 
\be
(\Delta  r)_{max} = R(1-1/\sin(\theta - {\Delta \theta \over 2}))\: .
\label{detar}
\en
Naturally this quantity is minimal for $\theta \simeq
90 ^o$ and diverges for $\theta \simeq 0^o$ when $\Delta \theta = 0^o
$ : the  error is minimal when the line of  sight is nearly tangential
($\theta  \simeq 90  ^o$) and  so almost  radial in  this  domain, and
maximal when  it is radial ($\theta =  0 ^o$). This in  principle is a
very bad  behavior, but the fact  is that the closer  $\theta$ is from
$0^o$ the weaker  the integrated perturbation is since  it gets always
more  degenerate   along  the  line  of  sight,   \ie  the  integrated
perturbations tend to a radial behavior and will therefore be absorbed
in the $\Phi_{DM,0}(r)$ term. The extreme situation, \ie when $\theta
= 0^o$  will trigger a mere  radial image as long  as the perturbation
exhibits a kind of axial  symmetry. This error is impossible
to alleviate  since we are  dealing with a fully  degenerate situation
but will not flaw the  method at all since the integrated perturbation
will be null. This approximation will be validated numerically below.

\subsection{How to obtain a X prediction ?}

The previously derived map offers a  great interest that we now aim at
exploiting, namely the ability of  precise X prediction. Indeed, for a
given  X  spectral  emissivity  model,  the  X-ray  spectral  surface
brightness   is  
\be   
S_X(E)   =  {1   \over   4\pi  (1+z)^4}   \int\ n_e^2\Lambda(E,  T_e)\
dl
\en  
where  $\Lambda$   is  the  spectral emissivity, $z$ is  the redshift
of the cluster and  $E$ is the energy on which the observed band is
centered. Hence we can write, assuming a satisfying knowledge of $z$
and $\Lambda$  :  
\bea 
S_X(E)  & \propto & \int\ n_e^2T_e^{1/2}\ dl \\ 
& \propto & \int\ \rho_g^2T_g^{1/2}\ dl \\
&  \propto & \int\ \rho_{g,0}^2 T_{g,0}^{1/2}\ dl + 2\ \varepsilon
\int\ \rho_{g,0}T_{g,0}^{1/2}\rho_{g,1} f(\theta,\varphi) dl
\nonumber\\  
&   &  +\: \: \: {1\over2}\ \varepsilon \int\ \rho_{g,0}^2
T_{g,0}^{-1/2} T_{g,1} f(\theta,\varphi) dl  
\ena 
where we omitted to write the  $(r)$s for clarity's sake. If we now
make use of the same approximation as  used and discussed  before, we
can express directly this quantity in  terms of observations $y$ and
$\phi$. We get indeed
\bea 
S_X(E) & \propto & \int \rho_{g,0}^2 T_{g,0}^{1/2} dl \nonumber\\
&  &  + \: \: \:2\ \varepsilon\ \rho_{g,0}(R)T_{g,0}^{1/2}(R) \int
\rho_{g,1} f(\theta,\varphi)  dl\nonumber \\ 
& & +\:  \: \: {1\over2}\ \varepsilon\ \rho_{g,0}^2(R)
T_{g,0}^{-1/2}(R) \ \int T_{g,1}f(\theta,\varphi) \\ 
& \propto  & \int \rho_{g,0}^2 T_{g,0}^{1/2} dl + 2\ \varepsilon\
\rho_{g,0}(R)T_{g,0}^{1/2}(R)\  D_{g,1}(R,\varphi) \nonumber\\  
& &  +\:  \: \:  {1\over2}\ \varepsilon\  \rho_{g,0}^2(R)
T_{g,0}^{-1/2}(R)  \zeta_{g,1}(R,\varphi) \:  .  
\label{x1}
\ena  
Both  the first order  terms   $T_{g,0}$  and  $\rho_{g,0}$,  and
the  second order corrections  $D_{g,1}$  and $\zeta_{g,1}$  have
been derived in  the previous sections.  We are thus able to generate 
self-consistently a X luminosity map from our previously derived
maps. This is a very nice feature of this method. We will further
discuss the approximation and its potential bias in the next section.\\

This derivation opens the possibility  of comparing on the one hand SZ
and  WL   observations  with,  on   the  other  hand,   precise  X-ray
measurements as done  \eg by XMM or CHANDRA. Note that  in the instrumental bands
of most of X-ray satellites the  $T_g$ dependence is very weak and can
be neglected. This can be easily taken into account by eliminating the
$T_g$ dependence in the previous formula. Even if the interest of such
a new comparison  is obvious we will discuss it  more carefully in the
two following sections.  In principle, one could also easily make some
predictions  concerning the  density weighted X-ray  temperature
defined by the ratio $\int n_g^2T_g^{}  dl / \int n_g^2  dl
\displaystyle$ but the fact is that since the gas pressure and so the
SZ effect tends to have a very weak  gradient we  are  not able  by
principle to  reproduce all the interesting features of this quantity,
namely the presence of shocks. 

\section{Application on simulations} \label{simulation}

\begin{figure*}[!tbh] 
\begin{center}
\hbox{
\psfig{figure = 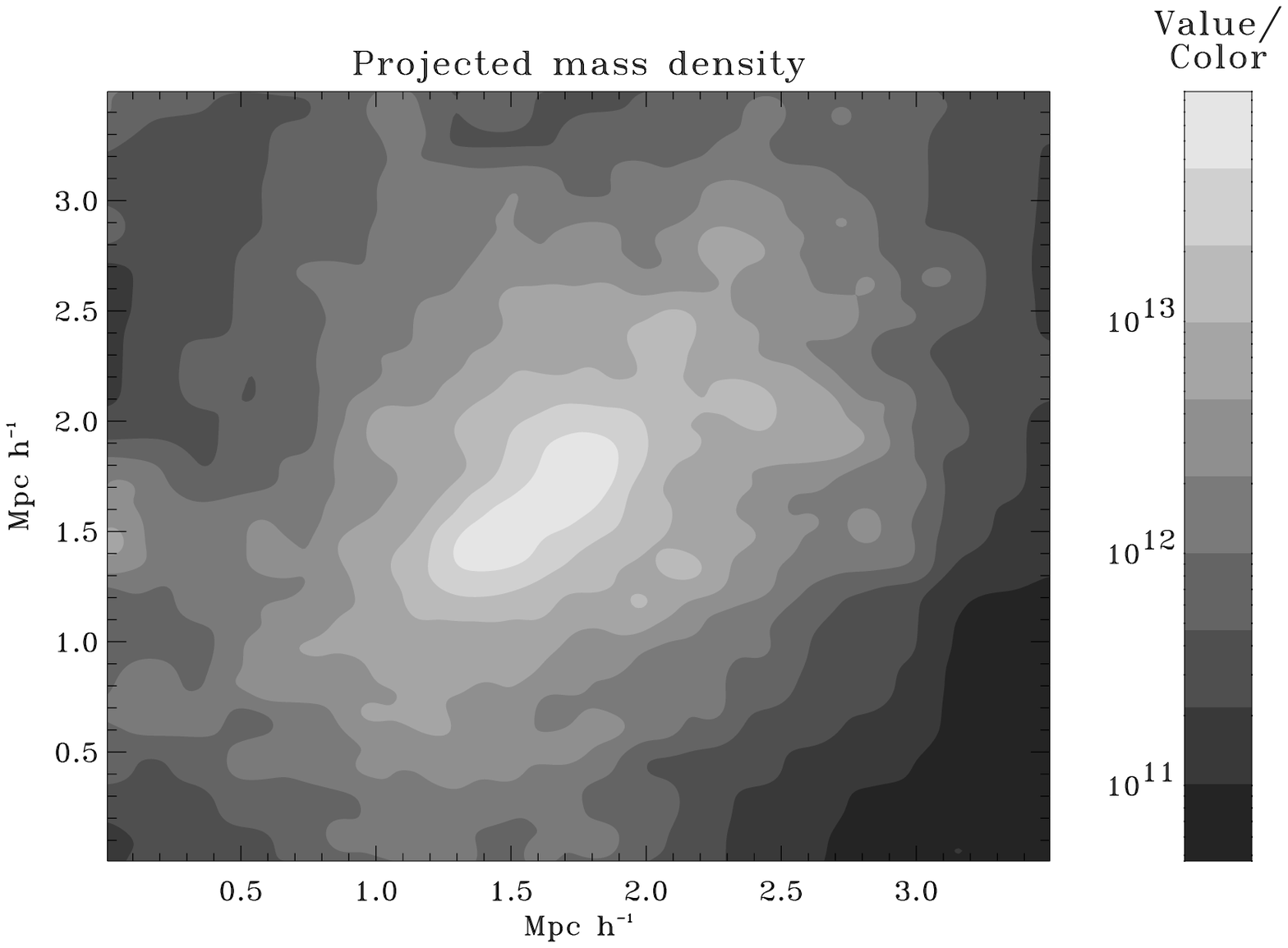,width=0.33\hsize,height=0.36\hsize,angle=0}
\psfig{figure = 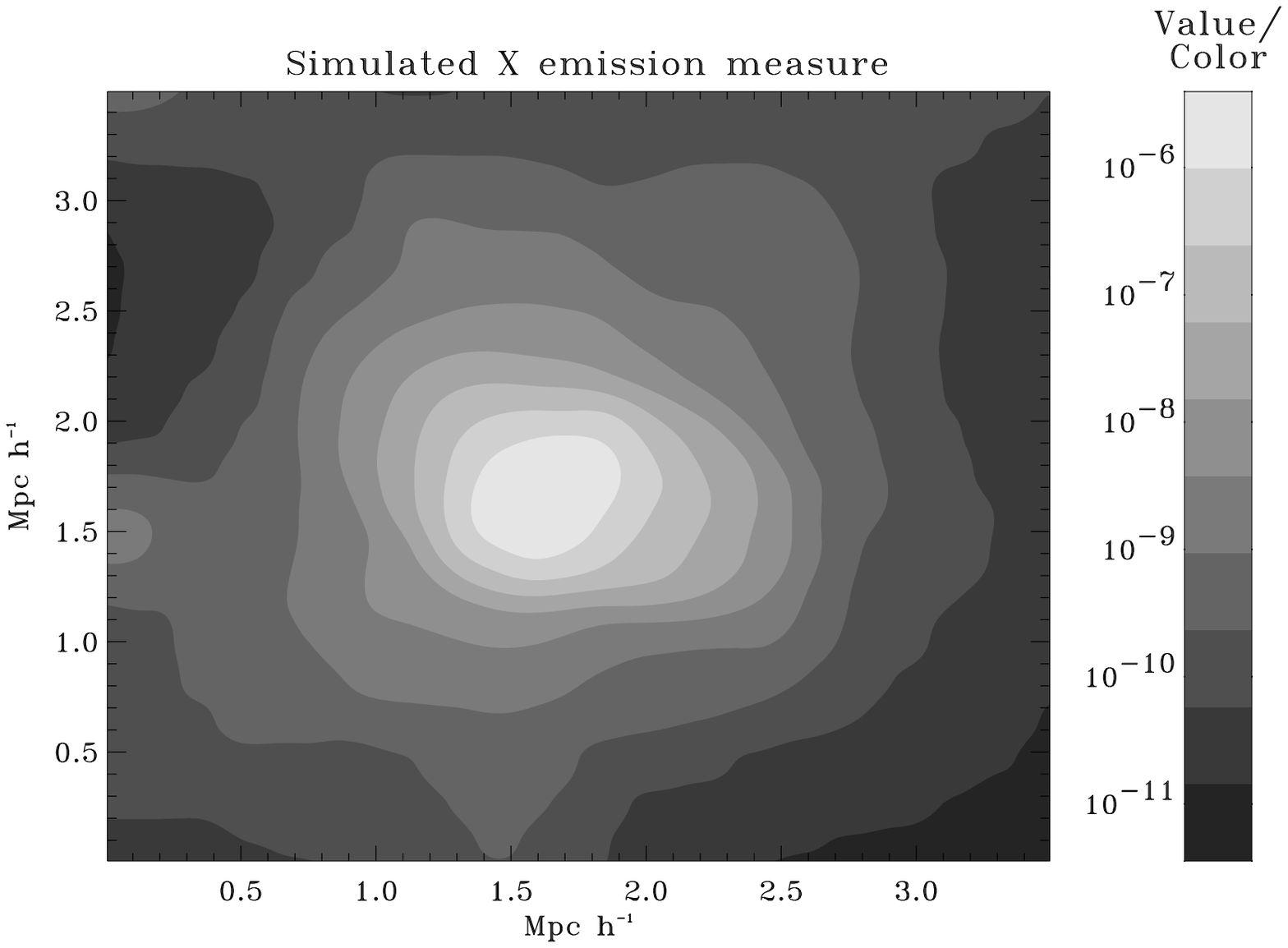,width=0.33\hsize,height=0.36\hsize,angle=0}
\psfig{figure = 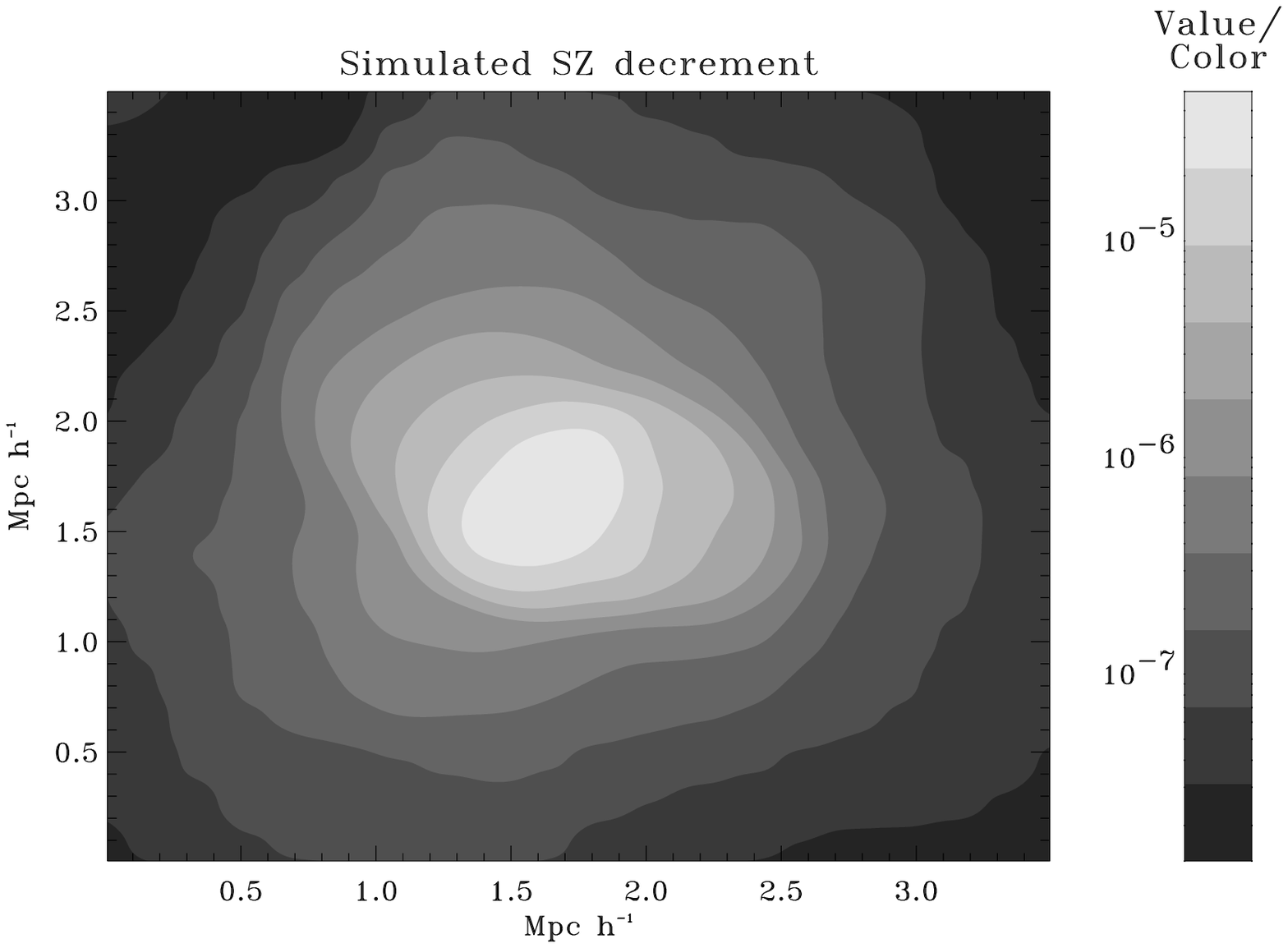,width=0.33\hsize,height=0.36\hsize,angle=0}}
\hbox{
\psfig{figure = 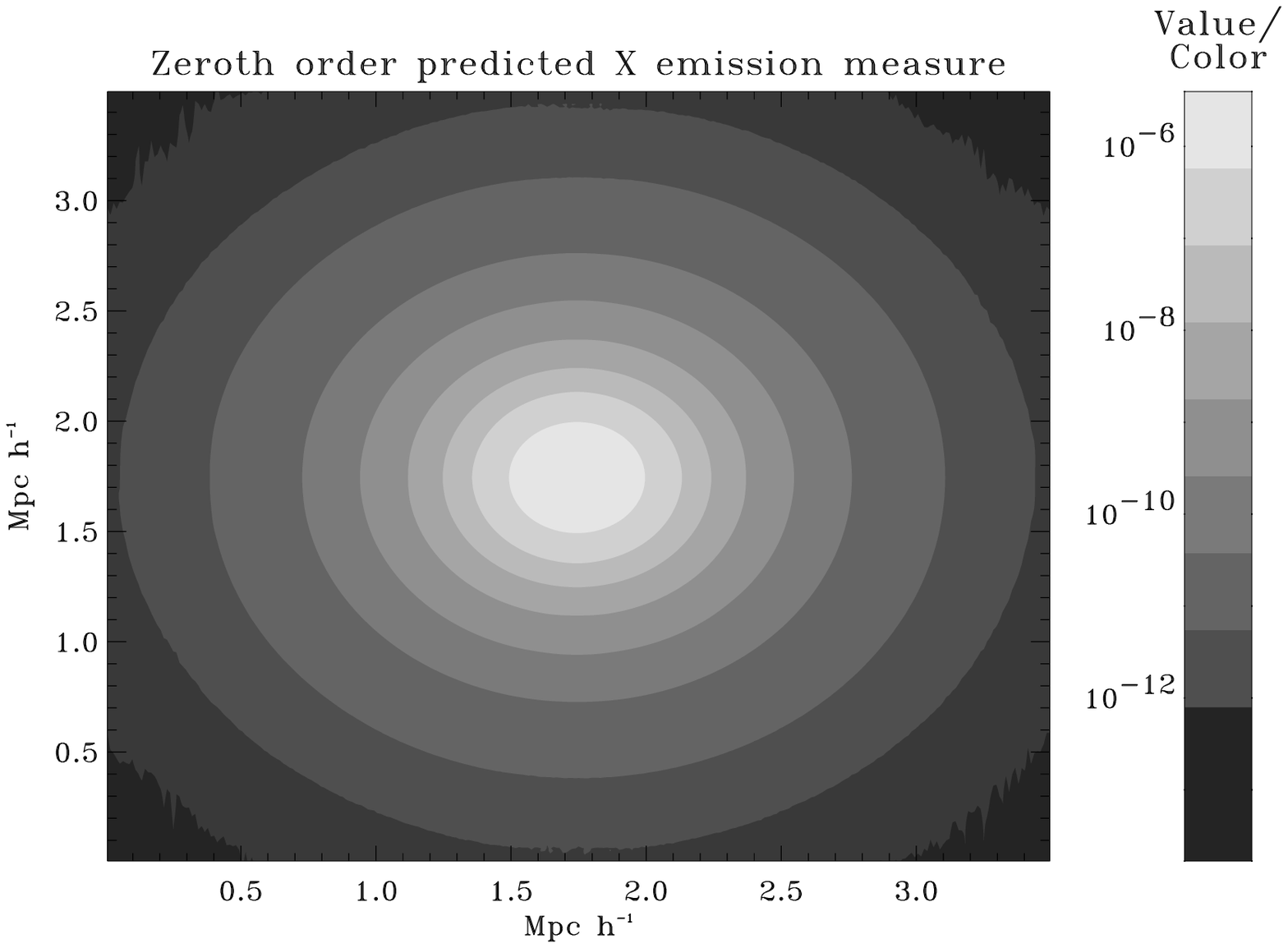,width=0.33\hsize,height=0.36\hsize,angle=0}
\psfig{figure = 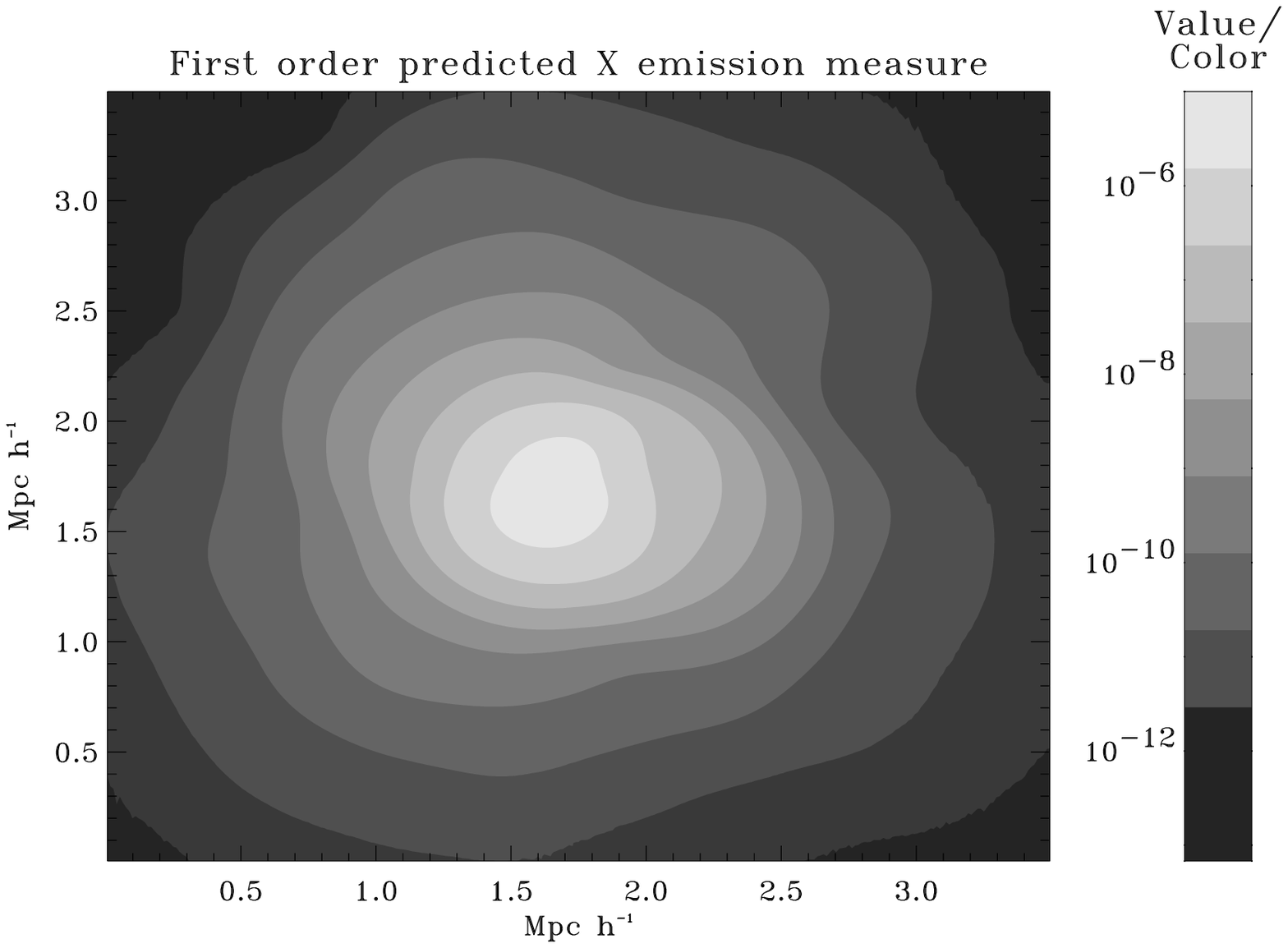,width=0.33\hsize,height=0.36\hsize,angle=0}
\psfig{figure = 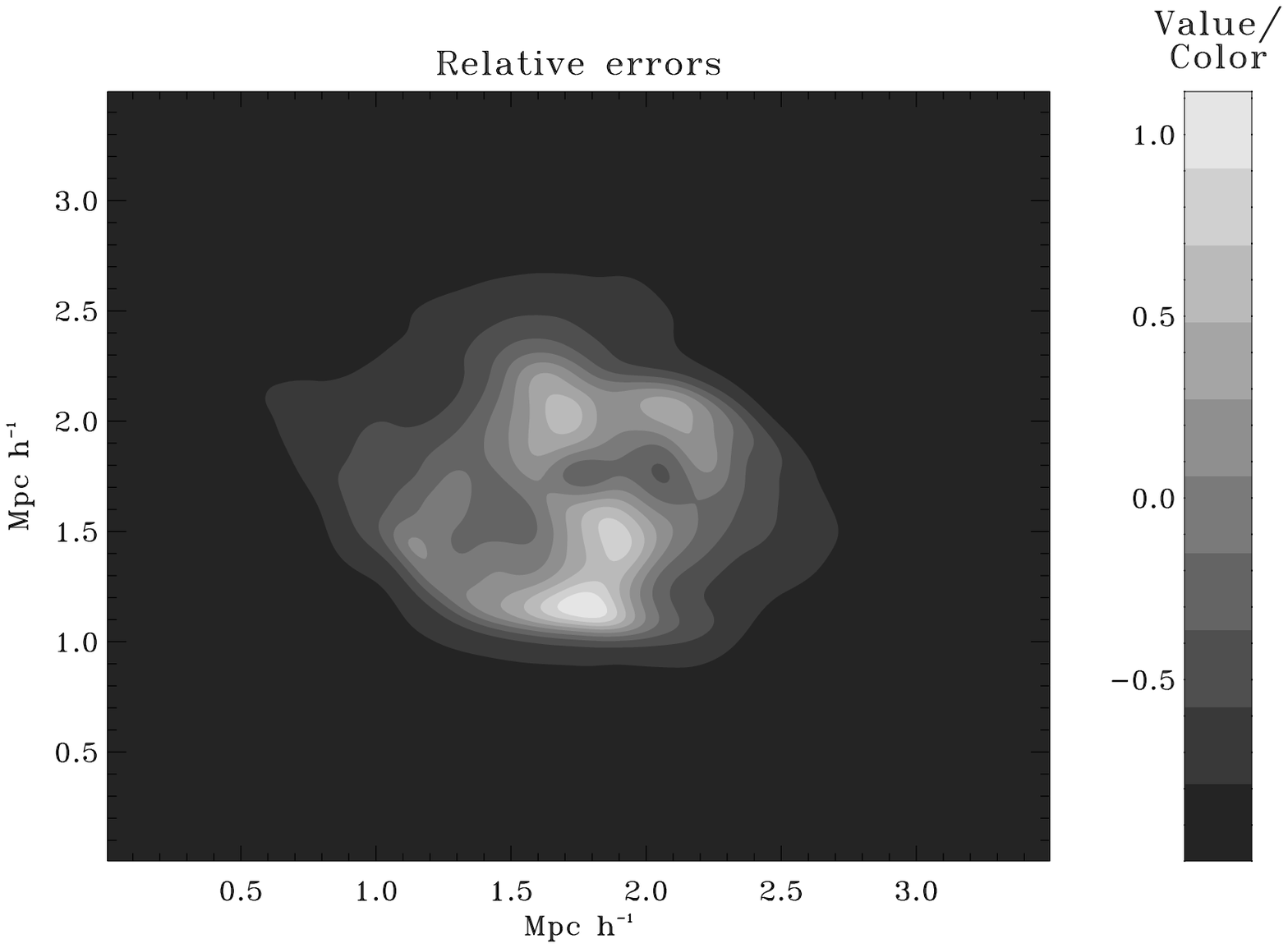,width=0.33\hsize,height=0.36\hsize,angle=0}}
\end{center}
\caption{ The upper panel shows the results of simulation, from left
to right, all using a logarithmic scaling, the projected mass density
($\mathrm{M_{\sun}\ Mpc^{-2}}$), the X-ray emission measure
($\mathrm{cm^{-6} Mpc}$) and the SZ $y$ parameter. This cluster is a
good candidate for our approach since it has a circular core with
surrounding perturbations so would be inadequate for a ellipsoidal
fit. The lower panel shows, from left to right a zeroth order
predicted X emission measure,  the first order prediction (the
zeroth order term plus the first order correction), both using a
logarithmic scaling   as well as well as the the relative error map,
\ie  (predicted - simulated)/simulated X emission measure
using a linear scaling. The 10 error contours are linearly
separated between -1.0 and 1. Each box is $3.5\ h^{-1} \mathrm{Mpc}$ wide. The
correlation coefficient between the predicted and the
simulated X-ray emission measure is $0.978\; $. The total flux differs
only by $0.91 \%$, thus even if the relative error map increases at
high $R$ the total error remains small due to the great dynamical
range involved. \label{cl1} }
\end{figure*}

\begin{figure*}[!bth] 
\begin{center}
\hbox{
\psfig{figure = 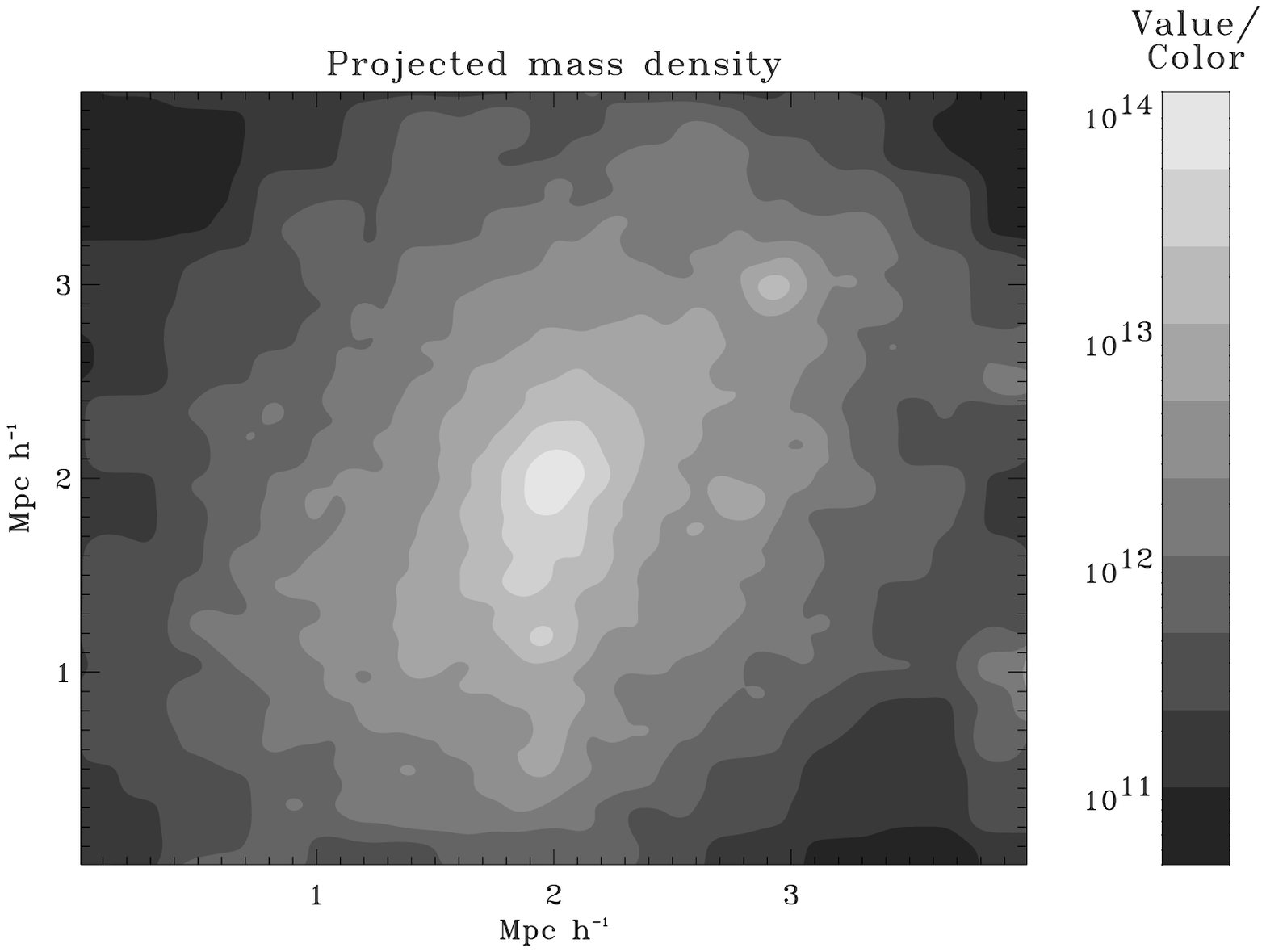,width=0.33\hsize,height=0.36\hsize,angle=0}
\psfig{figure = 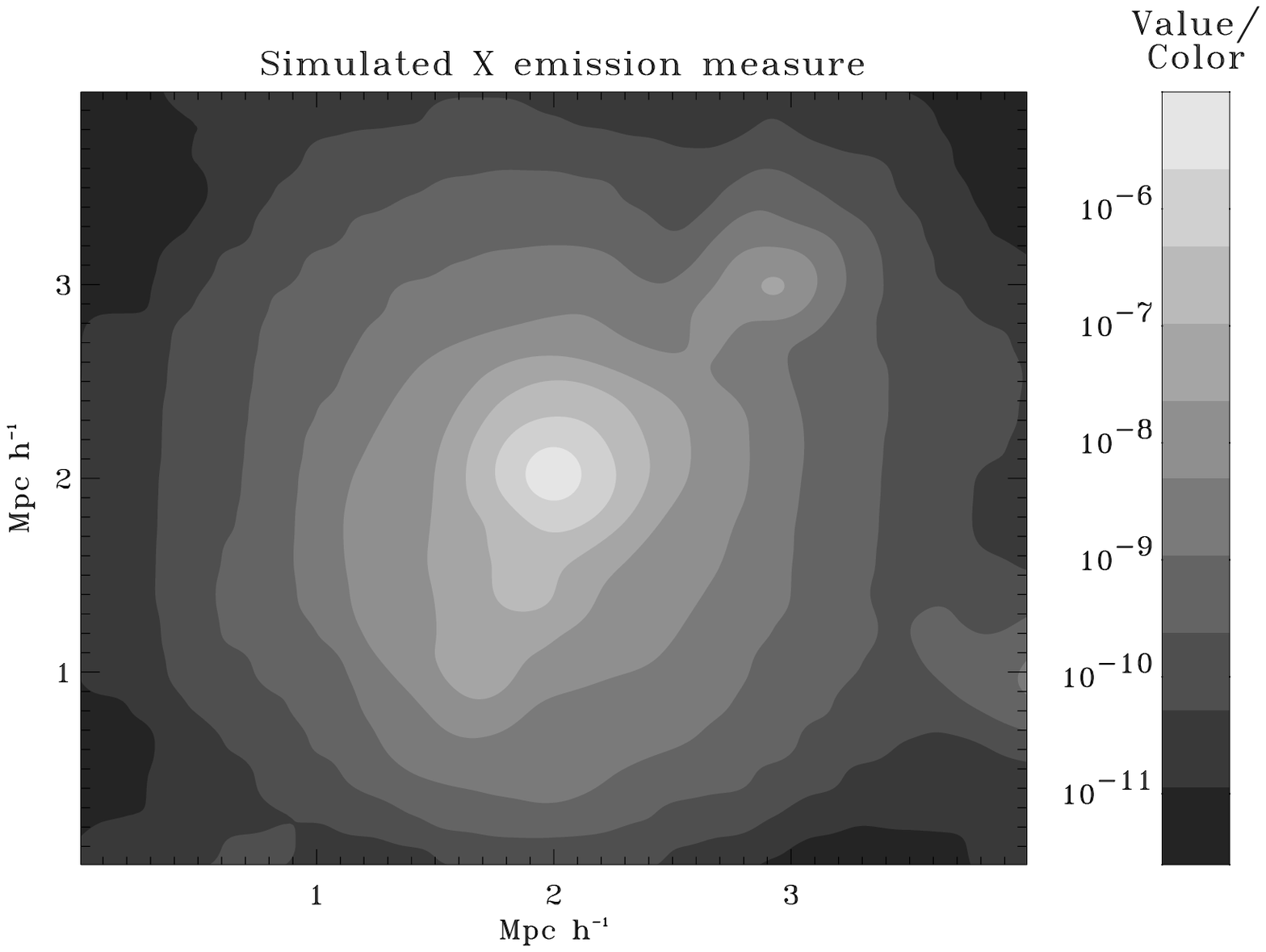,width=0.33\hsize,height=0.36\hsize,angle=0}
\psfig{figure = 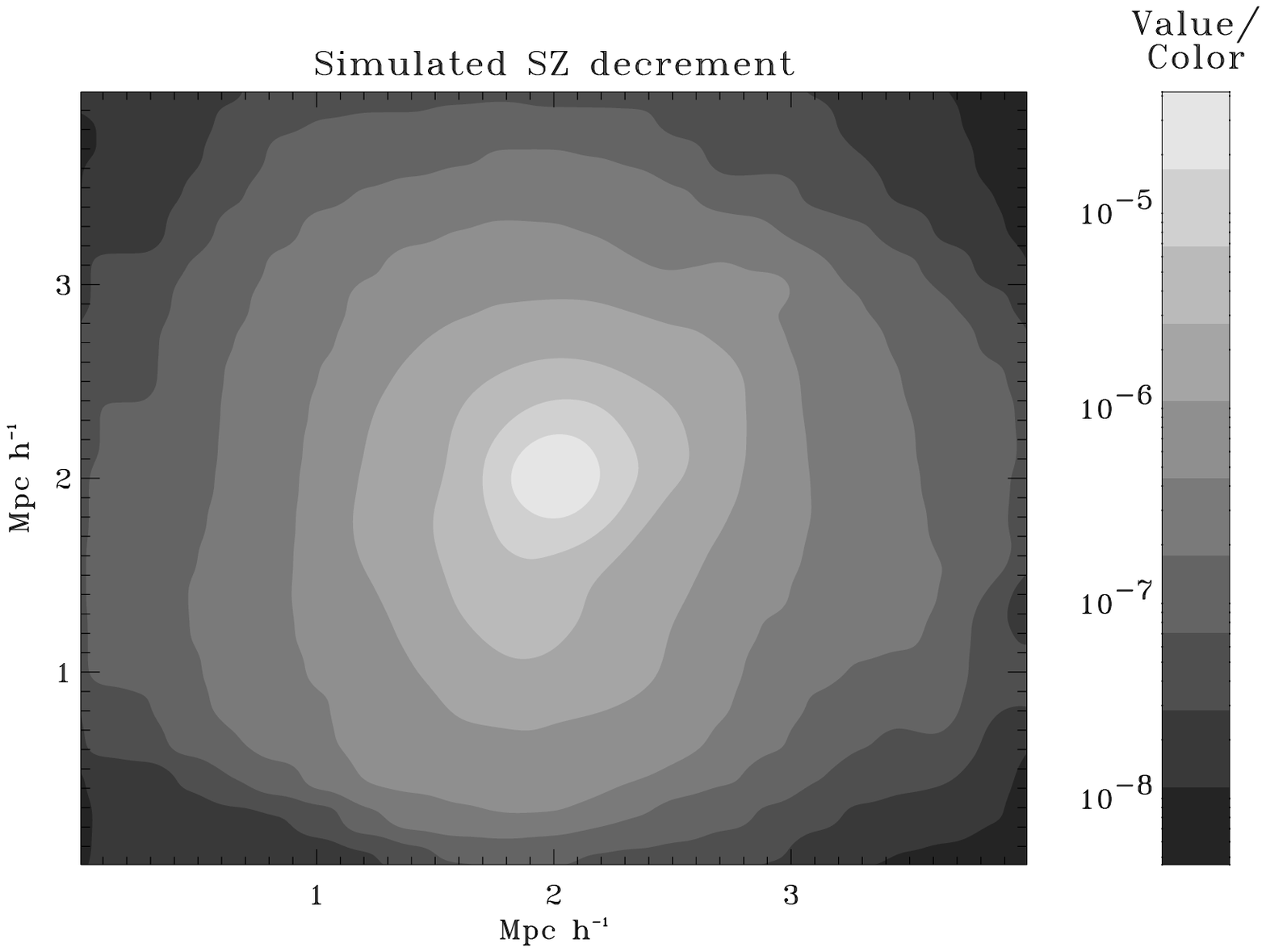,width=0.33\hsize,height=0.36\hsize,angle=0}}
\hbox{
\psfig{figure = 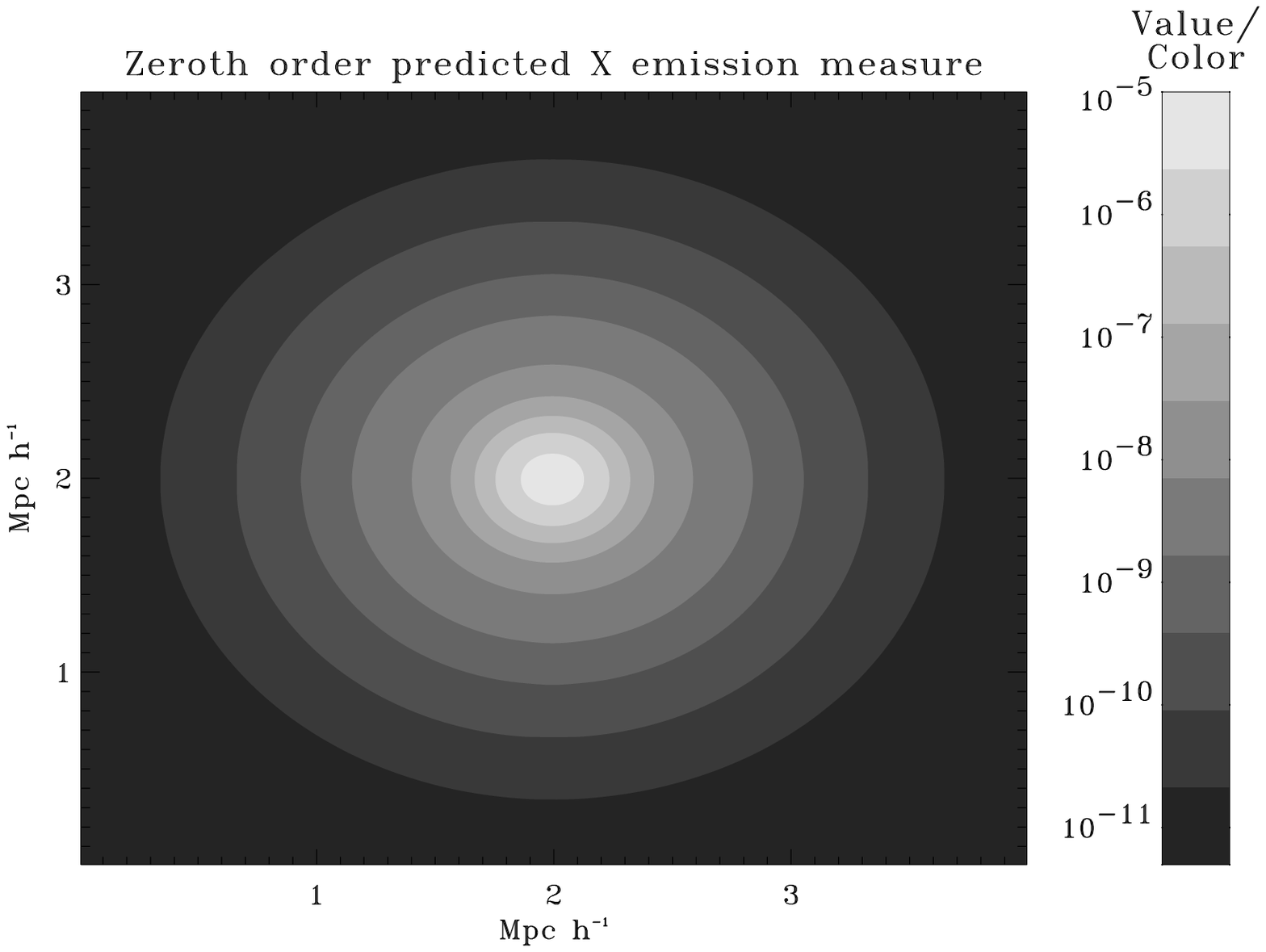,width=0.33\hsize,height=0.36\hsize,angle=0}
\psfig{figure = 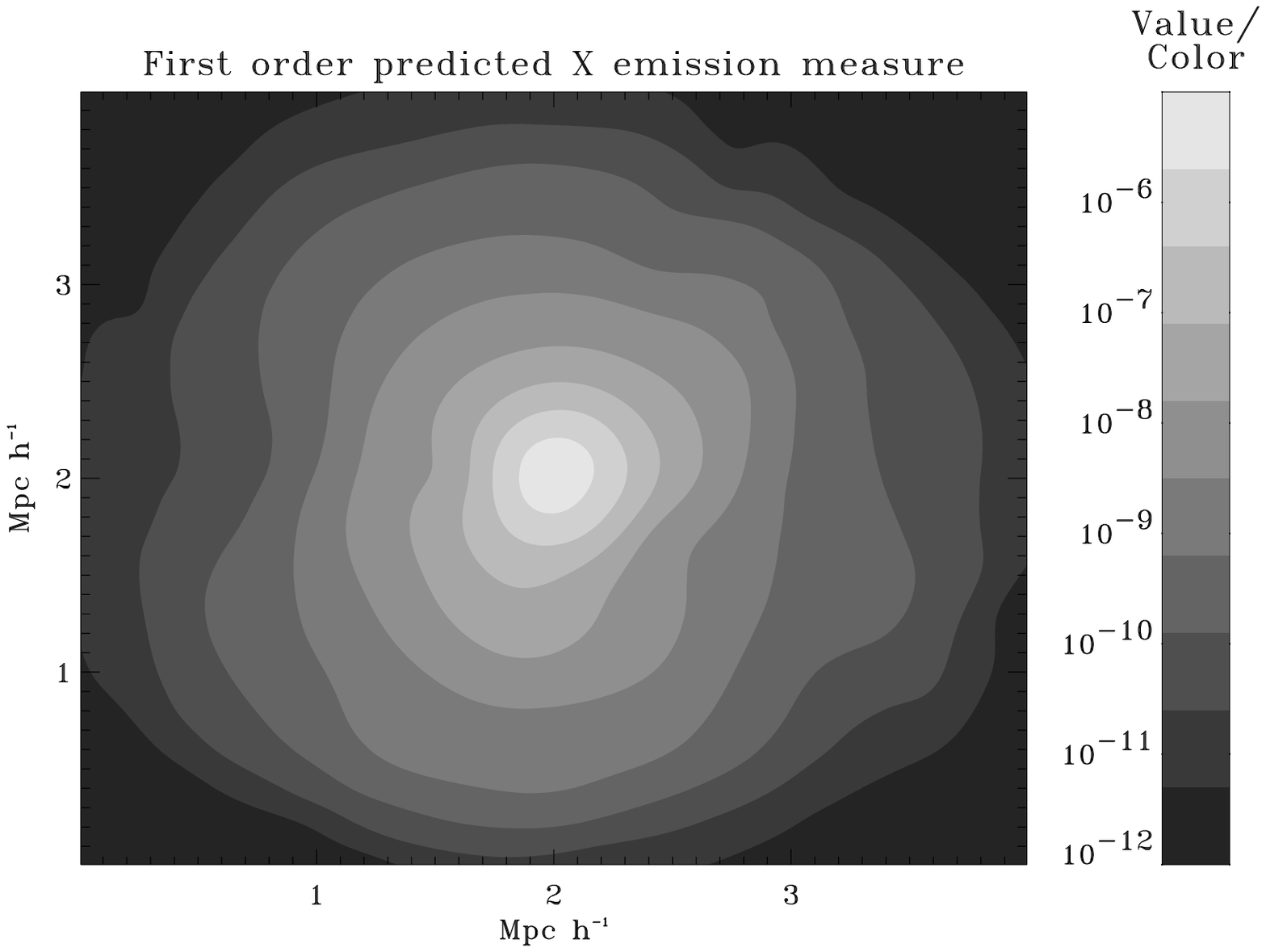,width=0.33\hsize,height=0.36\hsize,angle=0}
\psfig{figure =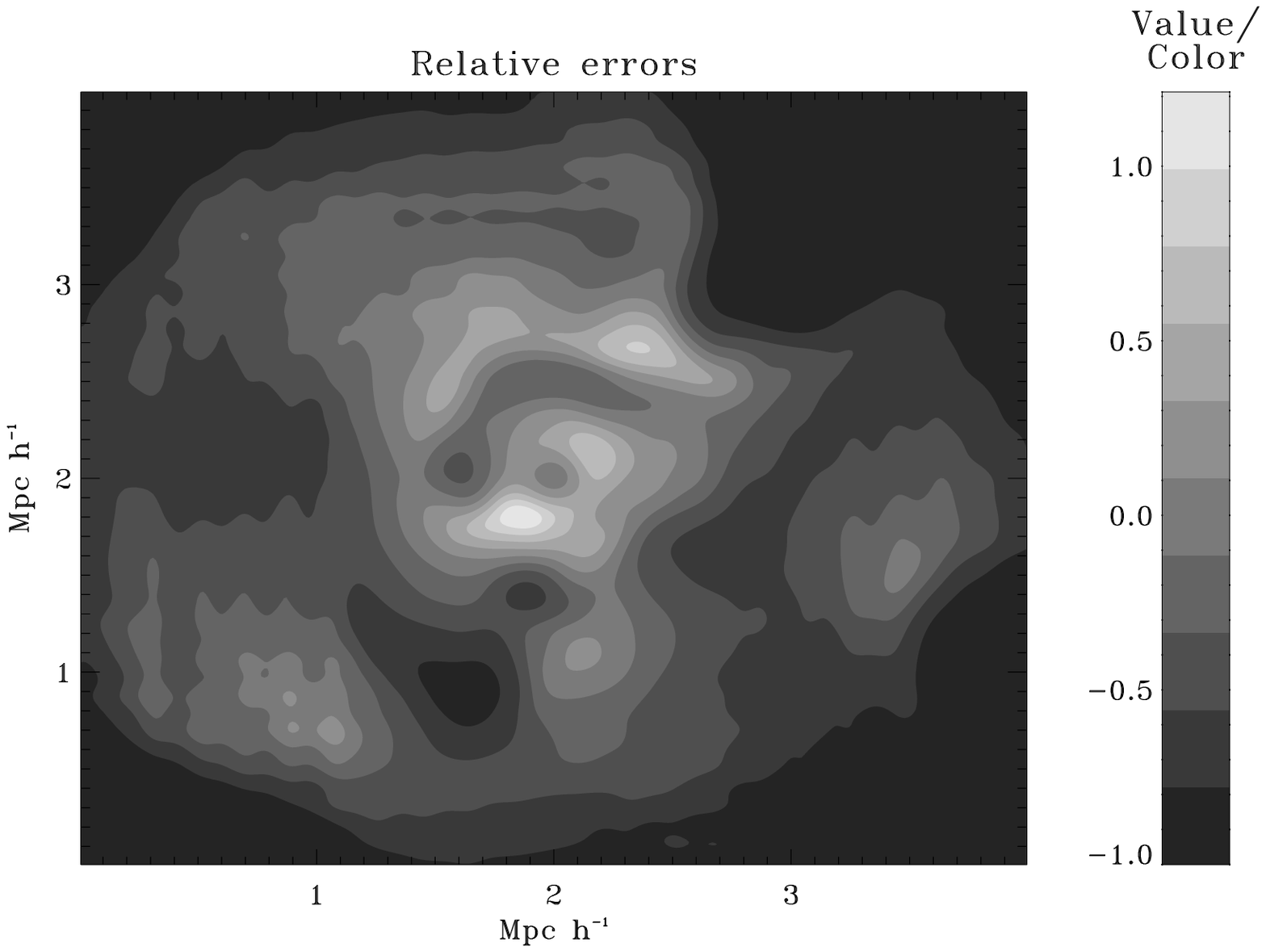,width=0.33\hsize,height=0.36\hsize,angle=0}}
\end{center}
\caption{ As the previous figure for a different cluster. The
structure of the X-ray emission measure is very well
reproduced for the inner part. The correlation coefficient between the
predicted and simulated map is $0.986\;$. As visible on the relative
error map, whose 10 levels are linearly separated between -1.0 and 1.0,
the outter part is naturally smeared by our approximation. The visible
1 o'clock  clump should be treated separately. Each box is $4.0\
h^{-1} \mathrm{Mpc}$ wide.  The total flux differ by $9\%$. \label{cl2}}
\end{figure*}

In order  to demonstrate the ability of the method in a simplified
context   we used some  outputs of  the recently
developed N-body + hydrodynamics code RAMSES simulating the evolution
of a  $\Lambda$-CDM universe. The  RAMSES code is based  on Adaptative
Mesh  Refinement  (AMR) technics  in  order  to  increase the  spatial
resolution locally using a tree of recursively nested cells of smaller
and  smaller size. It reaches  a  formal resolution  of $12\ \mathrm{kpc
h^{-1}}$ in the core of galaxy clusters (see Refregier and Teyssier
2000 and Teyssier  2001, {\it in preparation}, for details).  We use here  the structure of $2$
galaxy  cluster extracted  of the  simulation to  generate  our needed
observables,  \ie  X-ray   emission measure,  SZ  decrement  and  projected
density (or projected gravitational potential).

The relevant observables, \ie projected mass density, SZ decrement and
for comparison purpose only the X-ray emission measure,  of the 2
clusters are depicted using a logarithmic scaling in figure \ref{cl1} and
\ref{cl2} (upper panels). This clusters have been  extracted of the
simulation at $z = 0.0 $ and thus tends to be more relaxed. They are
ordinary clusters of virial mass (defined by $\delta_{334}$ in our
particular cosmology) $4.50\ 10^{14}\
h^{-1}\ \mathrm{M_{\sun}}$ and $4.15\ 10^{14}\ h^{-1}\
\mathrm{M_{\sun}}$.  Both exhibit rather regular shape, \ie they have
not undergo recently a major merge.  The depicted
boxes are respectively $3.5\ h^{-1} \mathrm{Mpc}$ and $4.0\ h^{-1} \mathrm{Mpc}$ wide. We smooth
the outputs using a gaussian of width  $120\ h^{-1}\mathrm{kpc}$ thus
degrading the  resolution. We did not introduce any instrumental
noise. This clusters are to a good approximation isothermal thus for
the sake of simplicity we will assume that $T_g$ is constant making
the discussion on  $T_{g,0}$ and $T_{g,1}$ useless at this point. We
apply the method  previously described using perturbed spherical
symmetry. We deduce by averaging over concentric annuli
a zeroth order circular description of the gas density and then
add to it some first order corrections. Note that since we assume
isothermality SZ data give us straightforwardly a projected gas
density modulo a temperature $T_{g,0}$ coefficient, thus we use the
formulation of equation (\ref{rho1y}), exact in this context. This
constant temperature is fixed using the hydrostatic equilibrium and the WL data.\\ 

In figure \ref{cl1} and \ref{cl2} (lower panels) we show the predicted
X-ray emission measure to zeroth and first order as well as a map
of relative errors. Note that to first order the shape of the emission
measure is very well reproduced. The cross-correlation coefficients
between the predicted and simulated X-ray emission measure are $0.978$
and $0.986$.   Of course this is partly due to the assumed good quality
of the assumed SZ data but nonetheless, it demonstrates the validity
of our perturbative approach as well as of our approximation. 
The approximation performed in equation (\ref{x1}), \ie the multiplication by the function
$\rho_{g,0}(R)$ will naturally tends to cut out the perturbations at
high $R$. This is the reason why the further perturbation are slightly
less well reproduced and the relative errors tend to increase with
$R$. Nevertheless, since the emission falls rapidly with
$R$ as visible on the lower figures (note the logarithmic scaling) the
total flux is well conserved, respectively to $0.9\ \%$ and $9\
\%$. This last number might illustrate that the large extent of the
perturbations in the second case may limit our method. An ellipsoidal
fit could have help decrease this value. Note that moreover the clump
visible mainly in X-ray emission measure  of figure \ref{cl2} is not
reproduce.  This is natural because it does not appear through the SZ
effect since the pression remains uniform throughout clumps. If
resolved by WL, this substructure should anyway be treated separately,
\eg by considering the addition of a second very small structure. Note
that the first cluster showed exhibits a spherical core elongated in
the outter region thus it is not actually as ellipsoidal as it looks
which may explain why our perturbed spherical symmetry works well.\\

\section{Discussion} \label{discussion}

\subsection{Hypothesis \ldots and non hypothesis}

Our  approach  makes  several  assumptions. Some  general  and  robust
hypothesis   have   been   introduced   and   discussed   in   section
\ref{hypo_gen}. Note that we do not need to assume isothermality. Our
key hypothesis consists in assuming the validity of a perturbative
approach and in the choice of the  nature of this  perturbations, \ie
with a radial/angular  part separation. Theoretical predictions, observations
and simulations  show that relaxed clusters are  regular and globally
spheroidal   objects,   which   is   what  initially   motivated   our
approach. Then in our demonstration  on simulations, this turns out to
be  reasonable.  Such  an  approach can  not  deal
properly  with   sharp  features  as  \eg  shocks waves due to  infalling
filaments.  Then  assuming the  validity  of  the  angular and  radial
separation,  leads to  the  equality  of this  angular  parts for  all
relevant physical quantities  ($P_g$, $T_g$, $\phi_{DM}$\ldots), using
to first  order in $\varepsilon$  the hydrostatic equilibrium  and the
equation of  state.  If this  is not satisfied  in practice  then we
could either question the validity of this separation  or the physics of
the cluster. Our experience with simulation shows that for reasonably relaxed
clusters, \ie not going through a major merge, the angular part of the
perturbation is  constant amongst observables. Thus it  looks like the
separation (and  thus the equality  of the angular perturbation) is a
good   hypothesis  in   general  and   its  failure   is  a   sign  of
non-relaxation, \ie non-validity of our general physical hypothesis.

Then an important hypothesis lies in the validity of the approximation
used.  Note first  that  even if  its  form is  general, its  validity
depends  on the quantity  which is  assumed to  be constant  along the
integral. In the case of the  gas density obtained from the SZ map, it
is an exact statement as soon as we assume the isothermality and since
clusters  in  general  are   not  too  far  from  isothermality,  this
hypothesis is reasonable. 

Now, some  worth to remember ``non hypothesis''  are the isothermality
and   the   sphericity  (or   ellipsoidality).   This   might  be   of
importance. Indeed,  in evaluating the  Hubble constant from  joint SZ
and X-ray  measurement it  has been evaluated  in \cite{InSu95,RoSt97,PuGr00}
that, both  the asphericity and the non-isothermality  of the relevant
cluster can yield  some important bias (up to $20  \%$).  Even if this
measure is not  our concern here, it is interesting  to note that this
hypothesis are not required here. 


\subsection{The equivalent spheroidal symmetry case}

So far, we have work and discussed the perturbed spherical symmetry
case. If we turn to spheroidal symmetry  the problem is very similar as long
as we  assume the knowledge of  the inclination angle  $i$ between the
polar axis of the system and the line of sight. This is what we recall
in appendix B which is directly inspired from \cite{FaRy84}: once the
projection is nicely parametrised we  get for the projected quantity ,
\eg  for  the   pressure  :  
\bea  
y(\eta)  &  =   &  2{B_e  \over  R} \int_{\eta}^{\infty} {P_{g,0}(t)\
tdt \over (t^2-\eta^2)^{1\over2}} \\
P_{g,0}(t)  &   =  &  -{1\over  2\pi}{  R   \over  B_e}  \int_t^\infty
P_{g,0}'(\eta)  \,{d\eta\over  \left  (\eta^2-t^2\right )^{1/2}}  \  .
\ena 
following the notations of appendix B. Since we are dealing with the
same Abel integral we can proceed in two steps as we did before.

Even if the  inclination  angle is \emph{a  priori}  not accessible  directly
through single observations it has been demonstrated that it is
possible to evaluate it using the deprojection of an axially symmetric
distribution of either X-ray/SZ maps or SZ/surface density  maps
\cite{ZaSq98,ZaSq00}. Our approach in this work try to avoid to
explicit the full 3-D structure rather than building it, and this is
done in a simple self-consistent way therefore we will not get into the details
of this procedure that will be discussed in a coming work (Dor\'e
\etal 2001, {\it in preparation}). Note also that axially symmetric
configuration elongated along the line of sight may appear as
spherical. This is a difficult bias to alleviate without any prior for
the profile. In our case, our method will be biased in the sense that
the deprojected profile will be wrong. Nevertheless, we might hope to
reproduce properly the global quantities, like abundance of DM or gas
and so to alleviate some well known systematics (see previous
section), \eg in measuring the baryon fraction.

\section{Conclusion and outlook}

It this paper we have presented and demonstrated the efficiency of an
original method allowing to perform in a self-consistent manner the joint analysis of SZ and WL
data. Using it on noise free simulation we demonstrated how well it
can be used to make some x-ray surface brightness prediction, or
equivalently emission measure. Our choice in this approach has been to
hide somehow the deprojection by using some appropriate
approximations. Thus we do not resolved fully the 3-D structure of
clusters,  but note that the work presented here is definitely a first
step towards a full deprojection (Dor\'e \etal 2001, {\it in
preparation}). Some further refinements of the  methods are under progress as well.\\ 

When applying the method to true data, the instrumental noise issue is
an important  matter of concern. Indeed, whereas  the strong advantage
of a parametric approach, \eg using a $\beta$-model, is that it allows
to  adjust the  relevant parameters,  \eg  $r_c$ and  $\beta$, on  the
projected  quantities (the image)  itself, which  is rather  robust to
noise, it might be delicate to determine the profiles and its derivate
by a direct deprojection.  Nevertheless, our perturbative approach, as
it first relies on a zeroth order quantity found by averaging over some
annulus, a  noise killing step (at least far from the center), and
then work on some mere projected perturbation should be quite robust
as well. Consequently we hope to apply it very soon on true
data. Furthermore, in this context it should allow  a better treatment
of systematics (asphericity, non isothermality,\ldots)  plaguing any
measure of the baryon fraction $f_b$ or the Hubble constant $\mathrm{H}_0$
using X-ray and SZ effect \cite{InSu95}. These points will be discussed
somewhere else (Dor\'e \etal 2001, {\it in preparation} ).

\section*{Acknowledgment}

O.D.  is grateful  to G.  Mamon,  M. Bartelmann, S. Zaroubi and
especially S. Dos Santos for valuable discussions. We thank
J. Calrstrom \etal for allowing the use of some of their SZ images.


\section*{Annexe~: Deprojection in spheroidal symmetry}

In  this appendix we  recall some  useful results  concerning spheroid
projection    derived   by    Fabricant,   Gorenstein    and   Rybicki
\cite{FaRy84}.  In  the   context  of  spheroidal  systems,  cartesian
coordinates system  are the most  convenient for projection.  Thus, if
the  observer's coordinate system  $(x,y,z)$ is  chosen such  that the
line of sight  is along the $z$  axis and such that the  polar axis of
the spheroidal system  $z'$ lies in the $x-z$  plane at an inclination
angle  $i$ to  the z-axis,  then, in  the cartesian  coordinate system
$(x',y',z')$ the  general physical quantities relevant  to our problem
depends  only  on  the  parameter  $t$  defined by  \bea  t^2  &  =  &
{x'^2+y'^2\over  B_e^2} +  {z'^2 \over  A_e^2} \\  & =  & {(x\cos  i +
yz\sin i)^2 + y^2 \over B_e^2} + {(z\cos i -x\sin i)^2 \over A_e^2}\:.
\ena If  we project a  physical quantity $  G(t)$ on the  observer sky
plane   $x-y$  then,   \bea  I(x,y)   &  =   &  I(\eta)   \\  &   =  &
\int_{-\infty}^{+\infty}\   G(t)  dl  \\   &  =   &  2{B_e   \over  R}
\int_{\eta}^{\infty}  {G(t)\  tdt  \over (t^2-\eta^2)^{1\over2}}  \ena
where  \bea  \eta^2 &  \equiv  & {x^2  \over  (RA_e)^2}  + {y^2  \over
(B_e)^2}  \\ \mbox{ and  } \displaystyle  R &  \equiv &  \sqrt{ {B_e^2
\over A_e^2}\cos^2i+sin^2i}\: .  \ena Of course this result shows that
if we were  to observe a spheroidal system we  would map ellipses with
an  axial ratio  equal  to $\displaystyle  {B  \over A}  = {1\over  R}
{B_e\over  A_e}$. But  the main  result of  this appendix  is  that we
obtain at the end an Abel  integral similar to the one obtained in the
case of  spherical system,  where the radius  as been replaced  by the
parameter $t$. This simple fact justifies the very analogous treatment
developed in this paper for spherical and spheroidal systems.


\end{document}